\documentclass{aa}  

\usepackage{graphicx}
\usepackage{txfonts}
\usepackage[
  breaklinks = true,
   colorlinks = true,
  urlcolor   = blue,
  citecolor  = blue,
  linkcolor  = red, ]{hyperref}
\graphicspath{ {./figures/} }

\usepackage{ulem}
\usepackage[rightcaption]{sidecap}

\newcommand{\logt}{log$(T$\,[K]$)$}

\newcommand{\kps}{km\,s$^{-1}$}

\usepackage{color}
\definecolor{dg}{rgb}{0.0, 0.5, 0.0}
\definecolor{applegreen}{rgb}{0.55, 0.71, 0.0}
\definecolor{dr}{rgb}{0.8, 0.2, 0.0}
\definecolor{orange}{rgb}{0.9, 0.5, 0.0}
\definecolor{db}{rgb}{0.0, 0.3, 0.8}

\begin{document} 

\title{Spectroscopic investigations of a filament reconnecting \\with coronal loops during a two-ribbon solar flare}

   \author{Reetika Joshi\inst{1,2,3,4}
          \and
          Jaroslav Dudík\inst{5}
          \and
          Brigitte Schmieder\inst{6,7,8}
          \and
          Guillaume Aulanier\inst{9,1}
          \and
          Ramesh Chandra\inst{10}
          }

     \institute{Rosseland Centre for Solar Physics, University of Oslo, P.O. Box 1029 Blindern, N-0315 Oslo, Norway\\
    \email{reetika.joshi@astro.uio.no}
    \and
    Institute of Theoretical Astrophysics, University of Oslo, P.O. Box 1029 Blindern, N-0315 Oslo, Norway    
    \and
    NASA Goddard Space Flight Center, Heliophysics Science Division, Code 671, 8800 Greenbelt Road, Greenbelt, MD 20771, USA
    \and
    Department of Physics and Astronomy, George Mason University, Fairfax, VA 22030, USA
         \and
     Astronomical Institute, Academy of Sciences of the Czech Republic, Fričova 298, 25165 Ondrějov, Czech Republic
     \and
    LIRA, Observatoire de Paris, Universit\'e PSL, CNRS, Sorbonne
    Universit\'e, Universit\'e de  Paris, 
    5 place Jules
    Janssen, F-92195 Meudon, France
    \and
    Centre for mathematical Plasma Astrophysics, Dept. of Mathematics, KU Leuven, 3001 Leuven, Belgium
     \and
    School of Physics and Astronomy, University of Glasgow, Glasgow G12 8QQ, UK
    \and
    Sorbonne Universit\'e, Ecole Polytechnique, Institut Polytechnique de Paris, Observatoire de Paris - PSL, CNRS, Laboratoire de physique des plasmas (LPP), 4 place Jussieu, F-75005 Paris, France
    \and
    Department of Physics, DSB Campus, Kumaun University, Nainital, India
    }


 
  \abstract
   {In the standard 2D model of eruption, the eruption of a magnetic flux rope is associated with magnetic reconnection occurring beneath it. However, in 3D, additional reconnection geometries are possible, in particular the $AR-RF$, where external reconnection involving the overlying arcades ($A$) and erupting flux rope ($R$) turns into another arcade and a flare loop ($F$). This process results in the drifting of the legs of the erupting flux rope.}
   {We investigated spectroscopic signatures of such $AR-RF$ reconnection occurring in an erupting filament reconnecting with coronal arcades during a weak B3.2-class two-ribbon flare.}
   {The evolution of the erupting filament eruption is examined using imaging observations  by the Atmospheric Imaging Assembly (AIA) as well as both imaging and spectroscopic observations by the Interface Region Imaging Spectrograph (IRIS).}
   {
   As the filament rises into the corona, it reconnects with the surrounding arcade of coronal loops with localized brightenings, resulting in the disappearance of the coronal loops and formation of a hot flux rope, showing slipping motion of its footpoints extending to the previous footpoints of the coronal loops ($AR-RF$ reconnection) as predicted by the 3D extensions to the Standard solar flare model. 
   These brightenings are accompanied by the presence of strong blue-shifts in both the IRIS \ion{Si}{IV} and \ion{Mg}{II} lines, up to $\approx$200\,\kps. The lines are also extremely wide, with non-thermal widths above 100\,\kps. Furthermore, a strongly non-Gaussian profile of the most blue-shifted component is detected at the start of the $AR-RF$ reconnection, indicating the presence of accelerated particles and MHD turbulence, and associated with the appearance of hot plasma in the AIA 94\,\AA~passband.
   }
  {
 For the first time, an observation has been reported in which the IRIS slit successfully captures $AR-RF$ reconnection between a filament and overlying arcades, resulting in strong blue-shifts and very broad line profiles. } 


   \maketitle
%

%
%
%
\section{Introduction}
\label{Sect:Introduction}
Solar eruptions are intense events, releasing a great amount of energy (10$^{32}$ ergs) stored in active regions (ARs). Photospheric motions in ARs lead to creation of twisted flux ropes which subsequently rise and later erupt due to instability \citep{Schmieder2015}. 
Theoretical models and observations indicate that the footpoints of flux ropes are surrounded by hook-shaped or J-shaped current ribbons \citep{Demoulin1996,Demoulin1997,Chandra2009, Pariat2012, Janvier2013,Janvier2014, Dudik_slipping_2014,Dudik2019, Aulanier2019, Joshi2024, Faber2025}. These current ribbons correspond to the intersections of quasi-separatrix layers (QSLs) at the photospheric boundary \citep{Zuccarello2017a}.
\subsection{2D and 3D reconnection geometries}
The classical 2D CSHKP solar flare model is based on the reconnection of the magnetic field lines under the erupting flux rope, leading to the growth of the flux rope as well as the creation of the closed flare loops, associated with heating in the chromosphere in two ribbons, where these loops are rooted. The growing flux rope is expelled as a coronal mass ejection (CME). In this classical picture, the reconnection happens in an X-type geometry, where bilateral outflows can be produced \citep{Priest2014}. Such outflows have been observed in exploding events or brightening during eruptions \citep[e.g.,][]{Innes1997,Cheng2015,Ruan2019,Joshi2022,Chen2024}.

This 2D standard model has been extended to the 3D \citep{Aulanier2010,Aulanier2012,Aulanier2019,Janvier2013}. \citet{Aulanier2019} found that in 3D, two additional reconnection geometries exist, both involving the erupting flux rope, $R$. 
These two additional reconnection geometries are: (1) $RR$--$RF$, where two flux-rope field lines ($R$-$R$) reconnect into another flux-rope field line ($R$) and a flare loop ($F$); and (2) $AR$--$RF$, where a coronal arcade ($A$) and a flux rope field line ($R$) reconnect and turn into a flux-rope ($R$) and a flare loop ($F$)). More specifically, in the $RR$--$RF$ reconnection, the magnetic field lines of the flux rope reconnect between them, in a manner similar to the reconnection of the overlying arcades in the 2D model, leading to a new highly twisted flux rope field line, and a flare loop. In addition to that, the erupting magnetic flux rope can reconnect with the surroundingcoronal arcades in an $AR$--$RF$ reconnection geometry. This latter geometry is purely three-dimensional \citep{Aulanier2019} and is connected to drifting of flux rope footpoints. 
The drift of the legs is connected to the evolution of the J-shaped flare ribbons. The inner edges of the ribbon hooks sweep outward and erode the flux rope from the inside. Meanwhile, the flux rope is built from the outside as the outer parts of the J-shaped ribbons sweep additional coronal arcades, which reconnect to become new flux rope field lines. As the simulation of \citet{Aulanier2019} is generic, the $AR$--$RF$ reconnection geometry is likely quite widespread in solar flares.

\begin{figure*}[t!]
  \centering
    \includegraphics[width=0.95\textwidth]{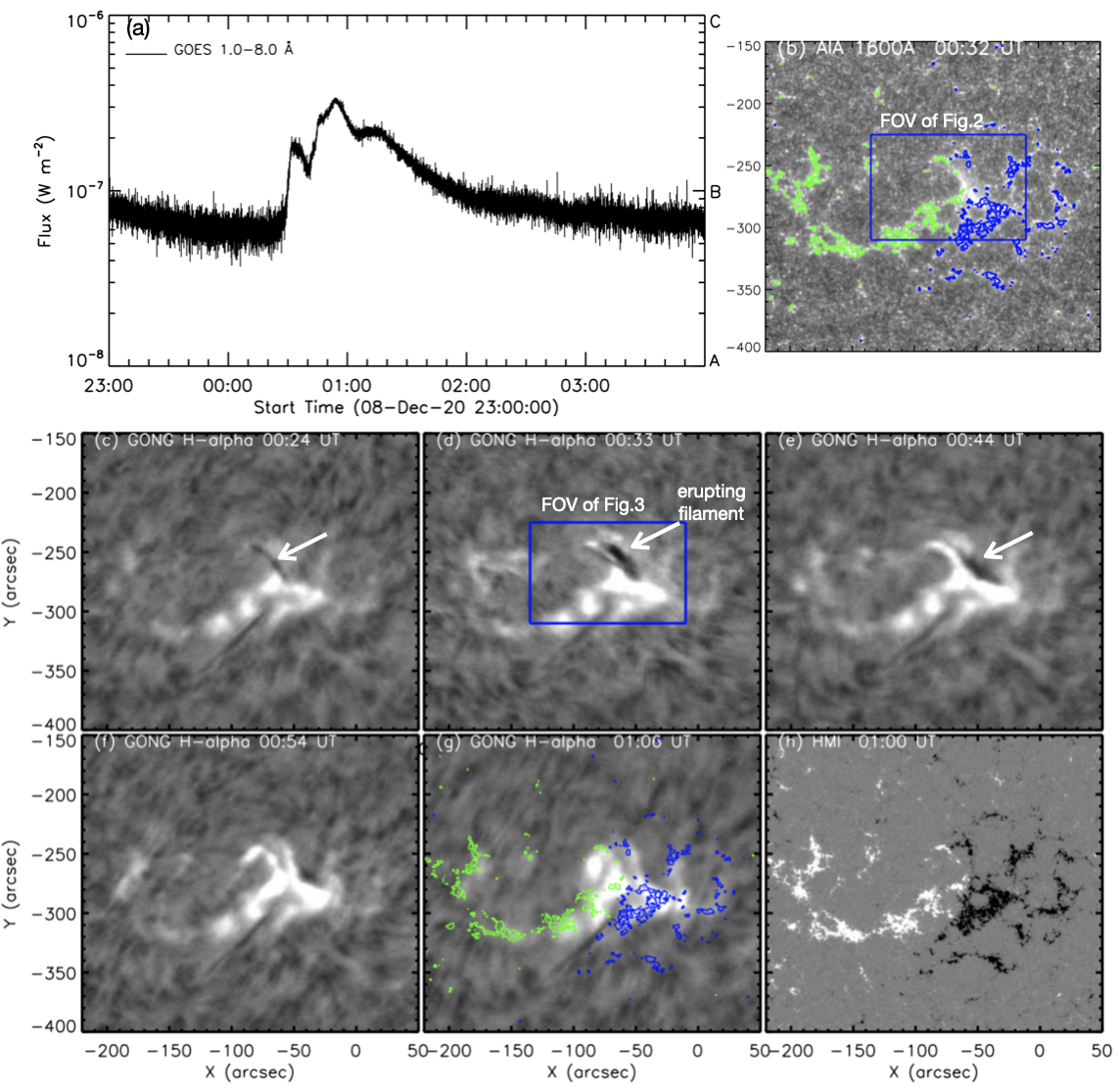} 
  \caption{GOES B3.2 class flare peaked at 00:53 UT, which has a small flare peak at 00:33 as a B1.7 class on December 9, 2020 (panel a). Panel b shows the flare ribbons in the UV temperature overplotted with magnetic field of strength $\pm$ 200 G (green for positive and blue for negative polarity) from HMI magnetogram. Panel c-g show the evolution of flare region in H$\alpha$ wavelength, where the filament is  shown with white arrows. Panel (h) shows the magnetic field configuration at the flare site and neighborhood. 
  }
 \label{Fig:GOES}
  \end{figure*}

\subsection{Observations of the Arcade-to-Rope (AR--RF) reconnection and the drift of the flux rope footpoints}

The existence of $AR$--$RF$ reconnection geometry was already shown by \citet{Aulanier2019}, where the J-shaped hooks of flare ribbons moved across the solar surface, first expanding and then contracting as predicted. \citet{Zemanova2019} reported drifting of flux rope footpoints via slipping motion of S-shaped loops associated with elongation and strong drift, by more than 40$\arcsec$, of the J-shaped ribbon. The new ribbon hook then again expanded and later contracted. \citet{Dudik2019} observed all four constituents of both the $AR$--$RF$ as well as $RR$--$RF$ reconnections, as well as the associated expansion and contraction of the ribbon hooks. \citet{Lorincik2021a} found that the flare loops resulting from $AR$--$RF$ reconnections, located at both ends of the flare arcade, are higher and more inclined, creating the characteristic saddle-shaped solar flare arcades. \citet{Lorincik2021b} found that the $AR$--$RF$ reconnections modify the extent of the coronal dimming areas inside the J-shaped ribbon hooks. 

More recently, \citet{Guo2023} quantified the rotation of the flux rope during its eruption and claimed that the drift of one footpoint of the flux rope  was due to reconnections of types \textit{AA--RF} and \textit{AR--RF}. The former injects magnetic fluxes and twisted field lines while the latter causes the displacements (continuous drift) of the flux rope footpoints. This effect can in some events be quite significant. For example, \citet{Gou2023} showed that the evolution of the J-shaped ribbon hooks can be so strong that the entire 
magnetic flux inside the erupting flux rope has been replaced during the course of an eruption. These observations indicate that the $AR$--$RF$ reconnections are likely widespread as significant for evolution of the erupting flux rope as predicted by \citet{Aulanier2019}. However, all of these observations so far relied on imaging datasets, which allow for tracking the evolution of individual emitting structures, but do not provide further information about the emitting plasma (with the exception of temperature). In particular, spectroscopic observations of magnetic reconnection in this geometry were so far missing.

\begin{figure*}[h!]
    \centering
    \includegraphics[width=\textwidth]{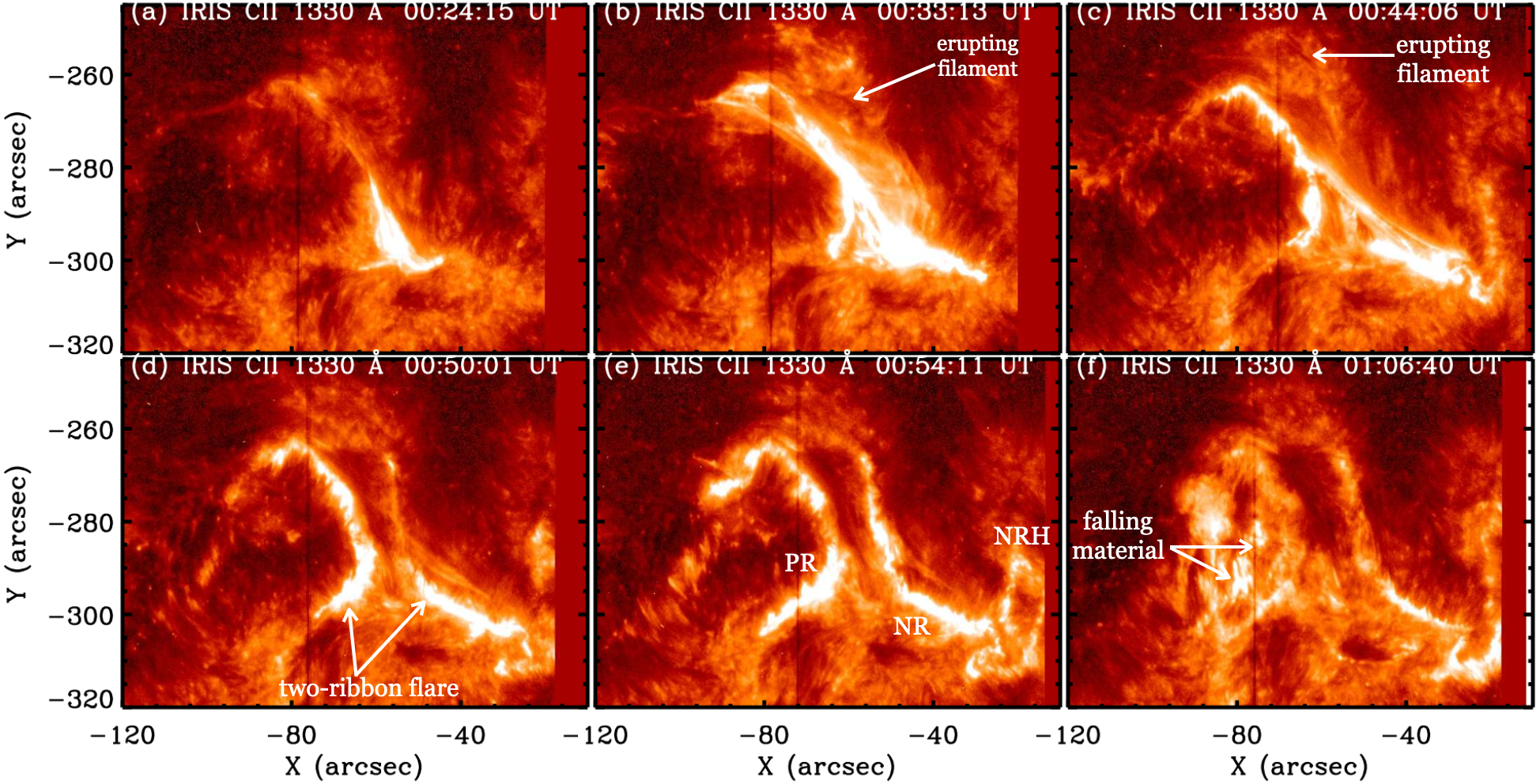}
    \caption{Evolution of flare ribbons and the filament eruption as observed in IRIS \ion{C}{ii} SJIs is depicted. Panels (b) and (c) highlight the erupting filament, with white arrows indicating the filament's increasing height over time. Panel (d) presents a clear view of the two ribbons forming in the flare region. In panel (e), the labeling of different ribbon regions is provided, as described in Sect.~\ref{Sect:Ribbons}. Finally, panel (f) illustrates the falling-back material during the later stages of the event.}
    \label{Fig:iris_sji}
\end{figure*}

 %
\begin{figure*}[h!]
   \centering
   
\includegraphics[width=\textwidth]{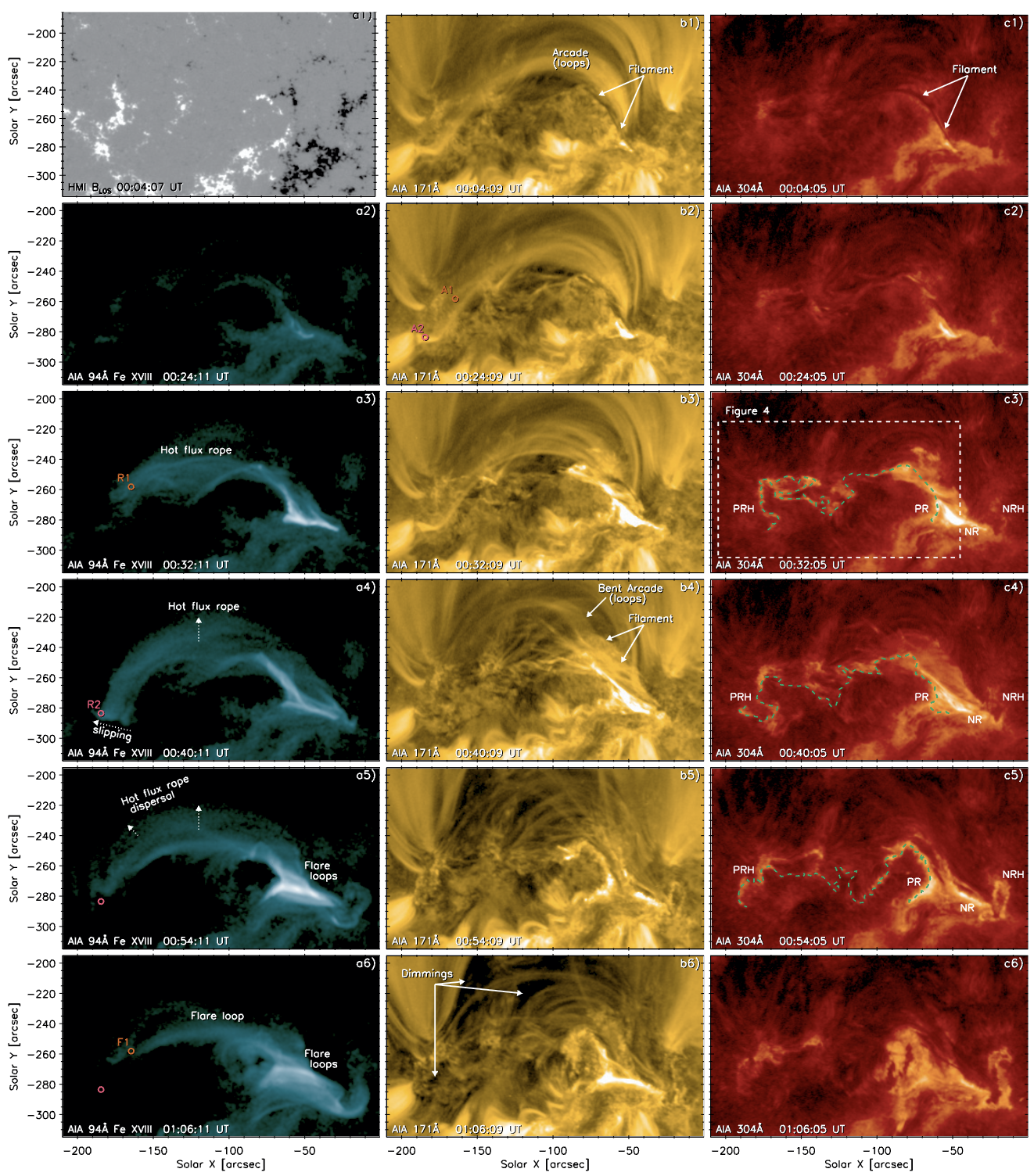}
\caption{Overview of the B-class flares and eruption of 2020 December 09 as observed by SDO in three AIA filters at 94\,\AA, 171\,\AA, and 304\,\AA. Note the \ion{Fe}{XVIII} contribution has been separated from the 94\,\AA~filtergrams (see Appendix \ref{Appendix:Fe18}.) The HMI magnetogram of a portion of the decaying AR is shown in the panel (a), and is saturated to $\pm$300\,G. One example of interaction between the filament with bent arcade (loops) is shown in panels (a4)--(b4) with the evolution of the hot flux indicated by white arrows in panel (a4). An online animation is attached with this figure (\url{https://drive.google.com/file/d/19Pyq0D3hgJgWYgUBpDNWhSZP-vh5n77-/view?usp=share_link}).}
\label{Fig:Overview}
\end{figure*}

Here, we report on the imaging and spectroscopic observations of a small B-class flare accompanied by a filament eruption that occurred on 2020 December 9 in NOAA AR 12791. 
The Interface Region Imaging Spectrograph \citep[IRIS;][]{Pontieu2014} captured very well the spectra of the reconnecting filament with 8 slit positions. 

%
%
%
\section{Observations}
\label{Sect:Observations}

Here, we summarize the imaging and spectroscopic observations used to analyze the $AR$--$RF$ reconnection in the B-class flare.

\subsection{SDO}
\label{Sect:Observations_SDO}

We use the observations from two instruments: the Atmospheric Imaging Assembly \citep[AIA;][]{Lemen2012}, and Helioseismic and Magnetic Imager \citep[HMI;][]{Scherrer2012} both instruments  aboard the Solar Dynamics Observatory \citep[SDO;][]{Pesnell2012}. AIA observes the full disk Sun and useful for studying the jet evolution in lower chromospheric or coronal temperatures. The AR was recorded with all EUV (94, 131, 171, 193, 211, 304, and 335 \AA) and UV (1600 and 1700 \AA) channels of AIA with cadence of 12 sec and 24 sec respectively. The pixel size of AIA is 0.$\arcsec$6 and the spatial resolution is 1$\farcs$5. HMI observes the photospheric magnetic field  with a cadence of 45 sec and its pixel size is 0$\farcs$5. We use these magnetograms for the AR evolution.

\subsection{IRIS}
\label{Sect:Observations_IRIS}

The AR  112791 was observed in coordination with IRIS spectrograph, which provides slit-jaw images (SJIs) and spectras between 00:24 UT and 01:16 UT. 
IRIS acquires spectra in three different ranges: far UV in C II  at 1334.54\,\AA~and 1335.72\,\AA~(FUV1), far UV in Si IV at 1393.76\,\AA~and 1402.77\,\AA~(FUV2) and near UV in Mg II at 2796.4\,\AA~and 2803.5\,\AA~(NUV). FUV1 and FUV2 are formed in upper chromosphere and transition region respectively. NUV mainly consists of the chromospheric Mg II h and k lines in addition with Mg II triplet lines formed at upper photosphere and lower chromosphere.  For this flare event, IRIS provides  large coarse  8-step raster with SJI in \ion{C}{ii} only with a cadence of 10 sec and FOV 119$\arcsec \times$ 119$\arcsec$,  also provides the spectra in \ion{C}{ii} 1330\,\AA, \ion{Si}{iv}  1402.77\,\AA, and \ion{Mg}{II} h and k lines with step size 2 $\arcsec$ and step cadence 9.8 sec, scanning a region of 14 $\arcsec \times 119 \arcsec$. The SJIs cover an extended region  ($ 119 \arcsec + 2 \times 14 \arcsec$), thus a rectangle of $148 \arcsec \times 119 \arcsec$.
We generated dust-free IRIS maps to clearly observe the evolution in IRIS SJIs.

\subsection{Data coalignment}
\label{Sect:Coalignment}

The level 1 data of the full-Sun AIA instrument traditionally has excellent pointing, although small offsets can still be are present in several channels, especially the longer-wavelength ones. To correct for these misalignment, as well as to study evolution of specific structures, we first corrected the data for differential rotation, choosing the time of 01:00\,UT as the reference one. Then, we found residual misalignment of $-0.9\arcsec$ in the Solar $Y$ coordinate in the 211\,\AA~and 335\,\AA~channels with respect to the other coronal channels (94\,\AA, 131\,\AA, 171\,\AA, and 193\,\AA), which do not show mutual misalignment, and whose pointing was thus assumed to be correct. 

The 304\,\AA~channel was found to have a misalignment of $\Delta X$\,=\,$-1.2\arcsec$, but no misalignment in $Y$. The 1600\,\AA~and 1700\,\AA~channels were found to have misalignments of $\Delta Y$\,=\,$-0.6\arcsec$ but no misalignment in the $X$ coordinate. All of these misalignments were corrected manually and we note that their uncertainties are typically $0.3-0.6\arcsec$ as larger shifts can readily be noticed in visual (blink) examination of morphologically similar channels, even though the resolution of the instrument is larger, $1.5\arcsec$. The HMI longitudinal magnetogram was aligned to the 1600\,\AA~channel of AIA, with no offsets being found. The resulting alignment being shown in panel (b) of Fig.~\ref{Fig:GOES}. 

Finally, the IRIS/SJI data were also manually aligned to SDO/AIA. To do that, we used the 1600\,\AA~channel of AIA, which shows many similarities to IRIS/SJI, such as plage polarities as well as flare ribbons. The IRIS data were found to have extremely good pointing, with only a shift of $-1.2\arcsec$ being necessary in Solar $Y$. We note that this is only a coordinate translation; that is, the observed IRIS spectra are not modified. The location of the IRIS slit with respect to AIA observations was determined using the Sobel edge-detection operator on the co-aligned SJI images. This works well for most SJI images, but failed in a few cases, where N--S oriented (slit-aligned) structures were present, such as continuous bright flare ribbons or large segments thereof. In these cases, the slit position was determined manually.

%
%
\begin{figure}
\includegraphics[width=0.5\textwidth]{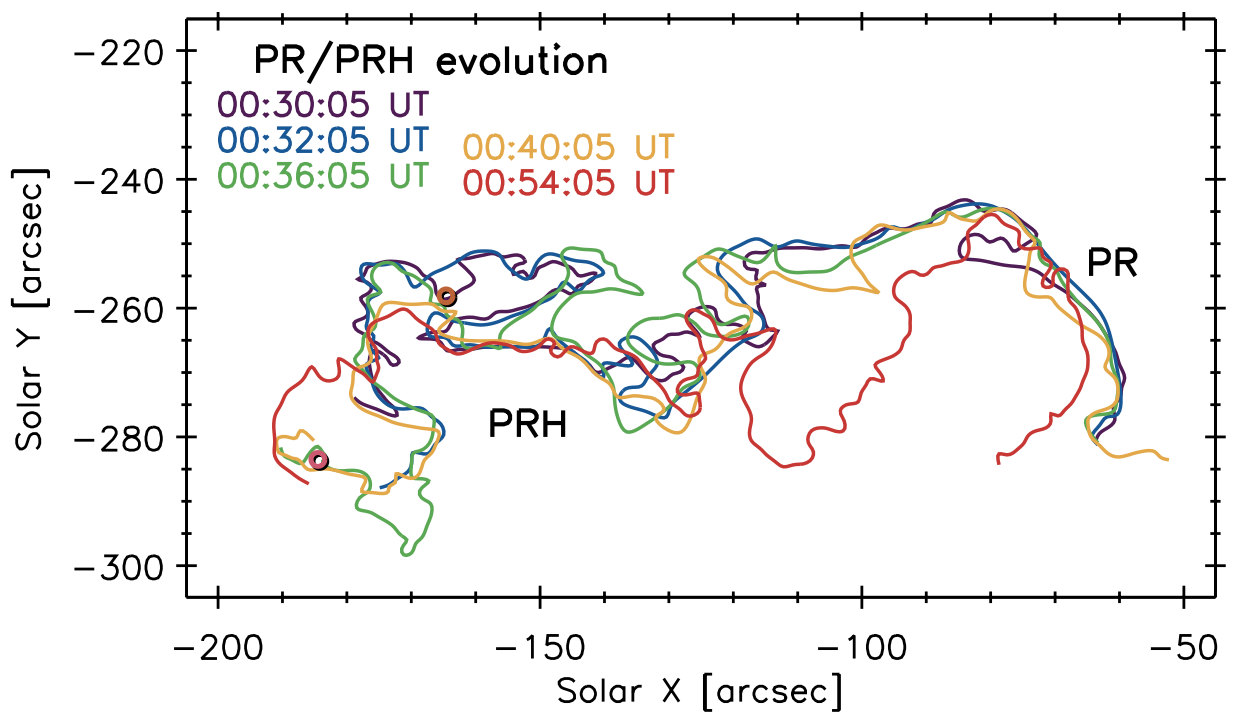}   
\caption{Evolution of the ribbon PR and its hook PRH, based on manual tracing of the ribbon. Individual colors correspond to individual times as indicated. Pink and orange circles indicate two spatial positions of interest. The orange one is outside of the PRH at 00:30\,UT, inside at 00:32\,UT, and again outside at 00:36\,UT. The pink one is first outside PRH, but is swept by the PRH at 00:36--00:40\,UT (see  text for details).}
\label{Fig:ribbon_evolution}
\end{figure}
%
%
\section{Evolution of the eruption}
\label{Sect:Evolution}

\subsection{Context} 
\label{Sect:Context}

The eruption occurred on 2020 December 9 in NOAA AR 12791 located at S15 W11. The GOES instrument registered two accompanying peaks of X-ray emission, at B1.7 at 00:33 UT, as well as at B3.2 at 00:53 UT. Both occurred during the course of the filament eruption (Fig.~\ref{Fig:GOES}). We believe that the small peak (B1.7) marks the onset of the B3.2-class flare, potentially indicating two distinct episodes of magnetic reconnection.
 The host AR 12791 is a decaying one, with a bipolar configuration consisting of only plage polarities. The following positive polarities are already quite dispersed and resemble a rather intense chromospheric network (Fig.~\ref{Fig:GOES} panels a-e). A small sunspot was still visible in the HMI continuum two days earlier in the leading negative polarities. A filament is detected in the GONG images along the inversion line from 23:00 UT on December 8 onward (Fig.~\ref{Fig:GOES}). 

In the coronal AIA channels, the AR shows a generally bipolar character, with a plethora of closed coronal loops as well as open fan loops at both eastern and western peripheries (see panel b1 of Fig. \ref{Fig:Overview} as well as Fig. \ref{Fig:CID_method}.) The filament eruption leads to significant evolution (reconnection) with the closed loops in the northern portion of the AR (see Sect.~\ref{Sect:AR-RF}).

\subsection{Evolution of flare ribbons}
\label{Sect:Ribbons}

The filament eruption is accompanied by two bright ribbons near the polarity inversion line. The development of the brightest parts of the ribbons near the filament is well captured by IRIS/SJI (Fig.~\ref{Fig:iris_sji}). At 00:24\,UT, an intense ribbon brightening appears, then extends along the inversion line at 00:33 UT (panels a--b). The erupting filament, parts of which are in emission, is located above this brightening (Fig.~\ref{Fig:iris_sji}b). At this time, it is not easy to distinguish both ribbons and the filament in the SJI image. Later, at 00:50\,UT, as the ribbons moved away from PIL, they can be more readily distinguished. Based on their respective polarities, the ribbons are denoted as PR (positive ribbon) and NR (negative ribbon), with PRH and NRH standing for their respective J-shaped hooks that encircle the footpoints of the erupting flux rope \citep[see][]{Aulanier2012,Janvier2014,Janvier2015,Aulanier2019}. The PR and its PRH are located in the eastern (left) part of the AR, with the PRH being quite elongated into the outlying dispersed positive polarities (see Figure \ref{Fig:Overview}, panels a1 and c3--c5). The NR and its much more compact NRH are located in the western (right) part of the AR. Therefore, the ribbons are quite asymmetrical.

As the filament eruption obscures the coronal loop footpoints in the vicinity of the NR and NRH, we studied the evolution of PRH in relation to the evolution of the erupting flux rope (Sect. \ref{Sect:Hot_channel}) as well as coronal loops and their footpoints (Sect. \ref{Sect:AR-RF}). The evolution of the PR and its hook PRH is rather complicated, as the hook is large and extended into the network area far away from the center of the decaying AR. We nevertheless managed to trace the ribbon PR and PRH in its entirety. This was a difficult task, as unique identification of a continuous ribbon in weak-field areas cannot be readily automated -- portions of the ribbon and especially its extended hook were found to not always be continuous at a similar intensity level, or in our weak B-class event significantly brighter compared to the scattered plage polarities. Identification of the ribbon in AIA 1600\,\AA~or ratio of 1600\,\AA~/~ 1700\,\AA~images was also not possible for these reasons. In addition, areas previously swept by the ribbon are known to remain bright for an extended period of time \citep{Lorincik2019a}. Furthermore, the ribbon can exhibit relatively fast squirming motions \citep{Dudik2016} as well as rapid elongations or even "jumps" to neighboring supergranules \citep{Lorincik2019a}. In our event, the squirming motions, as the ribbon searches for concentrations of magnetic flux to reconnect, were present especially in the weak-field areas. For these reasons, we resorted to manual tracing of the PR and PRH in AIA 304\,\AA~images, relying simultaneously on both the higher-resolution IRIS/SJI data as well as difference images of 304\,\AA~(cf., Appendix~\ref{Appendix:CID}) to identify both the ribbon and the direction of its motion.

The results of the tracing of the ribbon PR and its hook PRH are shown in 
Fig. \ref{Fig:Overview} as the green dashed line overlaid on the AIA 304\,\AA~observations (panels c3--c5). The evolution of the traced-out ribbon is also summarized in Fig.~\ref{Fig:ribbon_evolution}, showing the locations of the ribbon at five different times using different colors. It is seen that both the shape of the ribbons and its evolution is indeed not simple, as the ribbons contain many undulations related to the squirming motions. During some times, portions of the PRH stay approximately at the same place, as for example its far end at 00:32--00:40\,UT near the vicinity of Solar $X$\,=\,$-170\arcsec$ (blue, green, orange), while at other times, the ribbon or the end of its hook changed shape and moved by even several tens of arcsec (red). Although the shape of the PRH is complex, its interior is always well-defined, and is denoted by the 'PRH' label on Fig.~\ref{Fig:ribbon_evolution}. The relatively large change of the straight part of the ribbon at 00:54\,UT (red) near the inversion line is due to the closing of the PR as also observed by IRIS/SJI (panel f of Fig.~\ref{Fig:iris_sji}). This indicates ongoing erosion of the original filament flux rope and its gradual relocation to the far lobe of the PRH at Solar $X$\,$\lesssim$\,$-170\arcsec$.

%
\begin{figure*}
   \centering
    \includegraphics[width=\textwidth]{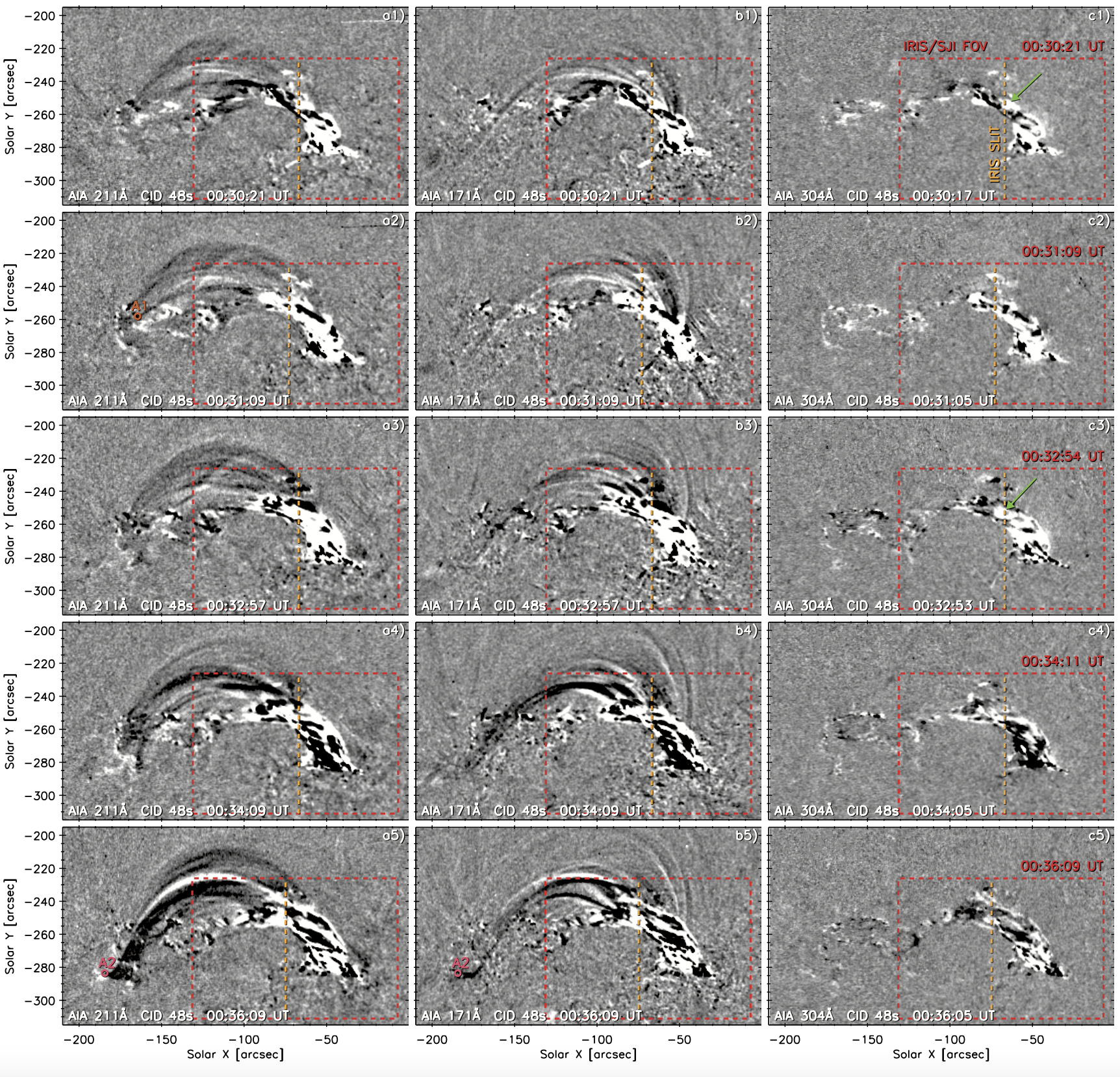}

\caption{Difference images highlighting the evolution of the filament and coronal loops during $AR$--$RF$ reconnection. The IRIS/SJI rectangular field of view is shown in red, and the position of the IRIS slit in orange. Panels (a1)--(c1), (a3)--(c3), and (a4)--(c4) show times corresponding to the IRIS spectra studied in Sect. \ref{Sect:Spectra}. Panels (a2)--(c2) and (a5)--(c5) are shown for context, indicating the footpoints of coronal loops (arcades) A1 and A2 that reconnected with the filament. At the respective times, the loops are dark structures, indicating that they no longer exist. Green arrows at 00:30:21 and 00:32:54\,UT indicate locations along IRIS slit where brightenings are observed within the filament. Animations of the CID images are available online (\url{https://drive.google.com/file/d/1kyI13cWtZFq5WNXYjTXxW9S3e_fJaZ5-/view?usp=sharing}).}
\label{Fig:CID_iris}
\end{figure*}
%


\begin{figure*}
   \centering
   \includegraphics[width=\textwidth]{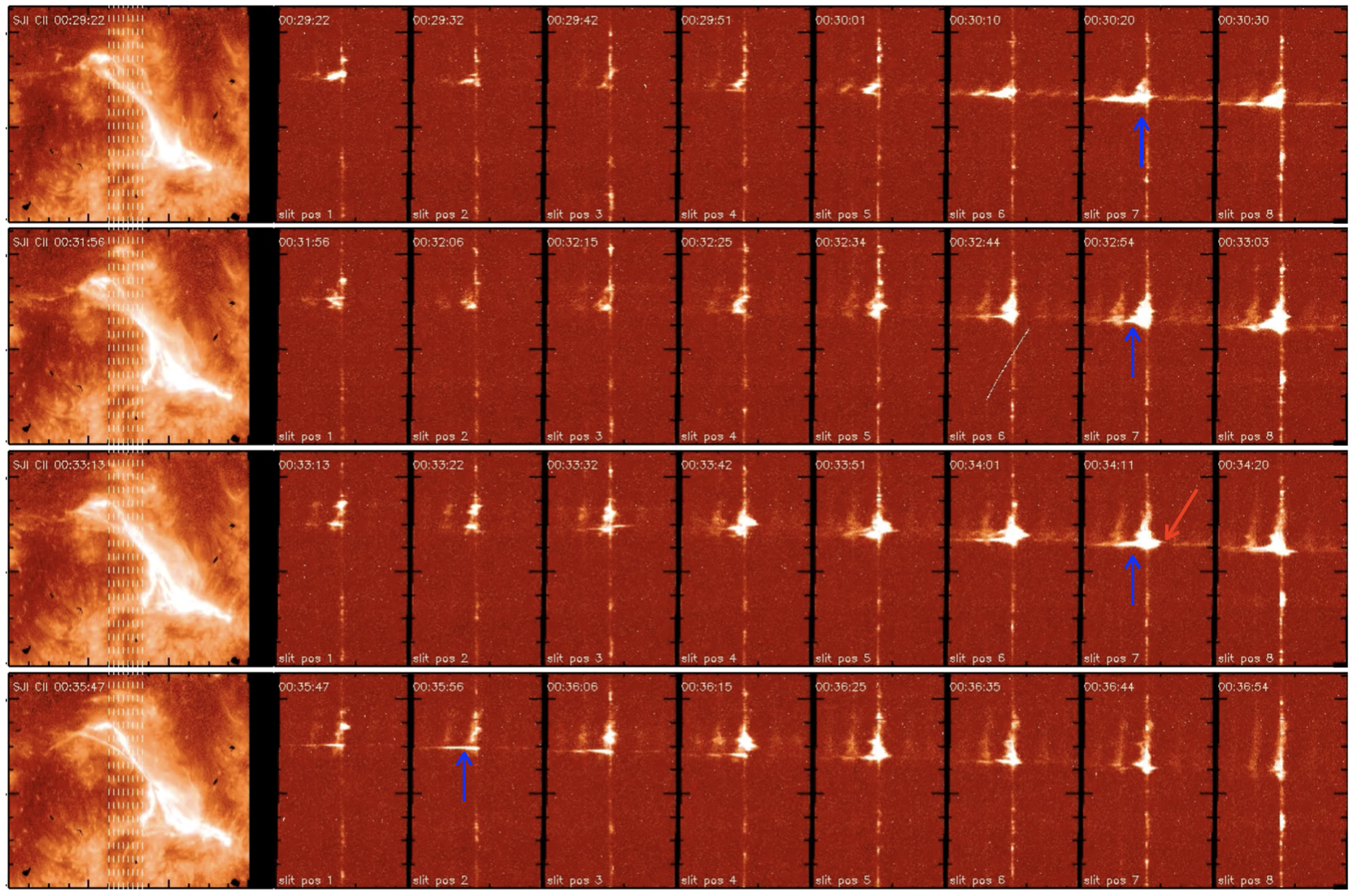}
   \caption{Overview of the \ion{Si}{IV} spectra of the reconnecting filament. Left panels show the C II SJI images, overlaid with all 8 slit positions as vertical dashed lines.
   Right panels show the corresponding 8 \ion{Si}{IV} spectra at each slit position. Prominent enhancements of the \ion{Si}{IV} 1402.8\,\AA~line, which can also be dopplershifted, are shown by blue and red arrows, respectively. Note the similar behavior of the \ion{O}{IV} line at 1401.2\,\AA. 
   An animation of this figure is available online (\url{https://drive.google.com/file/d/1UJAopX5rha0LOKc5W1YrseNEx09aawUb/view?usp=share_link}).
   }
    \label{Fig:sji_SiIVspectra}
    \end{figure*}
    
%
%
 \begin{figure*}
  \includegraphics[width=\textwidth]{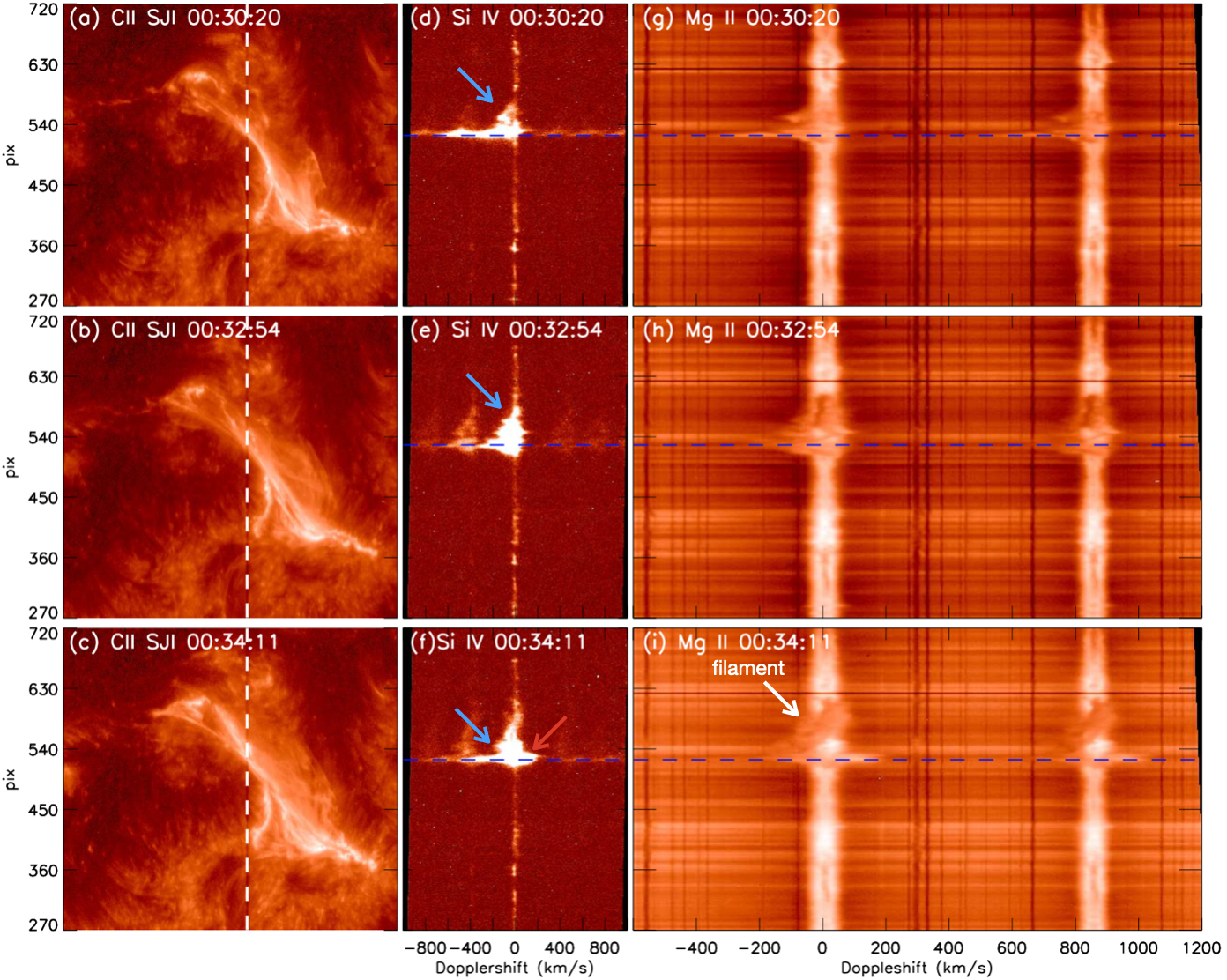} 
  \caption{Detailed \ion{Si}{IV} and \ion{Mg}{II} spectra of the three reconnection episodes detected at slit position 7 (see Fig. \ref{Fig:sji_SiIVspectra}). Panels (a)--(c) show the IRIS/SJI observations in \ion{C}{II} similarly as in Fig.~\ref{Fig:sji_SiIVspectra}, with the position of slit 7 is shown with a dashed white vertical line. Panels (d)--(f) show the \ion{Si}{IV} detector images, while panels (g)--(i) show the \ion{Mg}{II} ones at this slit position at three different times. The dashed horizontal blue line indicates the reconnecting filament spectra with large dopplershifts plotted in Fig. \ref{Fig:spectra_}. 
  The blue/red arrows in panels d-e-f indicate the blueshift and redshift flows during the reconnection in Si IV spectra. The white  arrow indicates the filament in panel(i).}
\label{Fig:mgii}
\end{figure*}

%
%
\begin{figure*}[!t]
  \centering
\includegraphics[width=0.9\textwidth]{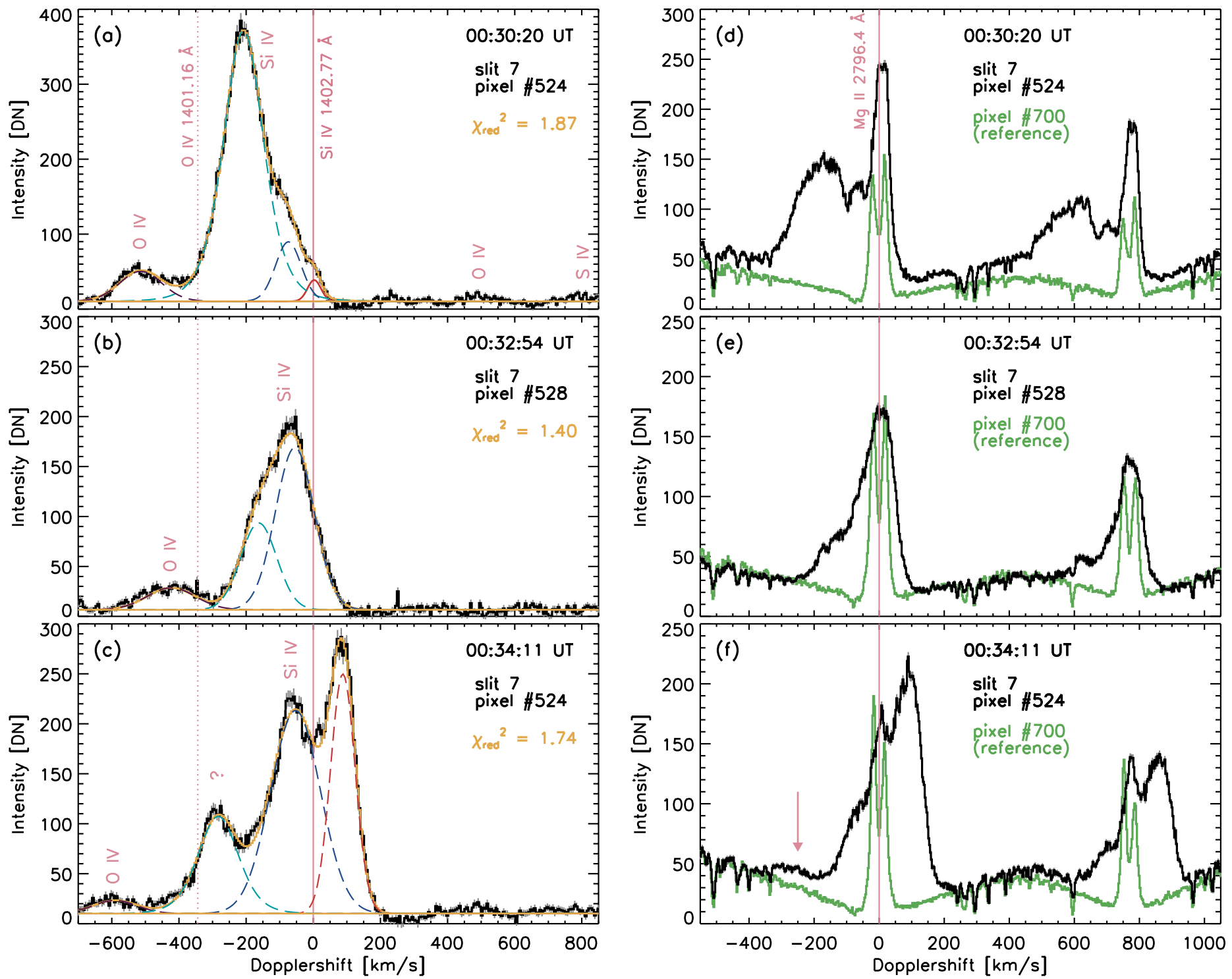}  
    \caption{Observed IRIS spectra (black) of \ion{Si}{IV} (a--c) and \ion{Mg}{II} (d--f) at three different times as shown in previous figures (horizontal blue dashed lines in Fig.~\ref{Fig:mgii}). Uncertainties of the observed spectra are denoted by thin gray error-bars. 
    The fit to the \ion{Si}{IV} spectra is shown in orange, with individual components indicated by colored lines (blue and red for blue- and red-shifted components of the \ion{Si}{IV} line, violet for \ion{O}{IV} line). Reference spectra for \ion{Mg}{II} are shown in green, and are taken in the non-flaring chromosphere to the north of the erupting filament. Vertical pink lines denote rest wavelengths for individual spectral lines. At 00:34:11 UT there are several components in both lines, in particular one is corresponding to 250 km~s$^{-1}$ (pink arrow in panel f and the "?" sign in panel c). See text for details.}
\label{Fig:spectra_}
\end{figure*}

\subsection{Evolution of the erupting flux rope}
\label{Sect:Hot_channel}

The evolution of the PRH is accompanied by the evolution of the erupting flux rope itself. Overall, the eruption starts as an erupting filament that later reconnects with coronal loops, producing a hot flux rope \citep["hot channel", see, e.g.,][]{ZhangJ2012,Zhang2023,Cheng2013,Cheng2014,Cheng2015,Song2015,Dudik2016,Hernandedz-Perez2019,Liu2022b,Zhang2023}. In our event, the hot channel is best detected in the AIA 94\,\AA~channel (see Fig. \ref{Fig:Overview}~a2--a5). Given that the GOES flare is rather weak, reaching only B3.2 level, the hot channel is correspondingly rather weak, having only $\approx$5--10 DN\,s$^{-1}$\,px$^{-1}$ in AIA 94\,\AA. Given that the 94\,\AA~channel of AIA is multithermal \citep[see][]{ODwyer2010,DelZanna2013multith}, containing contributions from \ion{Fe}{XVIII} formed around \logt\,=\,6.8 in solar flare conditions, as well as much cooler "coronal" contributions from \ion{Fe}{X} and \ion{Fe}{XIV}, we employed the \ion{Fe}{XVIII} separation method of \citet{DelZanna2013multith} to remove contributions other than the eruption-generated hot emission. In employing this method, we took the degradation of AIA sensitivity into account (Appendix \ref{Appendix:Fe18}).

Overall, the event is an eruptive one, producing a pair of J-shaped ribbons, with rising (erupting) filament and later a growing hot channel that shows slipping motion of its footpoints outward along its hooks around 00:40\,UT (Fig.~\ref{Fig:Overview}a4, and the accompanying animation). Subsequently, the hot channel grows and disperses outwards in the direction of the original filament eruption. This evolution of the hot channel is denoted by the white dotted arrows in panels a4--a5 of Fig. \ref{Fig:Overview}. We call this evolution `dispersal' rather than `eruption', because despite the evolution being consistent with an eruption, there is strong decrease of the signal in the AIA 94\,\AA~channel after 00:54\,UT. No trace of the erupting hot channel is left after 01:06\,UT (panel a6). At this time, well into the decay phase of the B3.2 flare (Fig. \ref{Fig:GOES}a), only a gradually cooling flare loops are left in the location of the eruption. 

Finally, the event is accompanied by progressive and pronounced coronal dimmings (see panels b5--b6 of Fig. \ref{Fig:Overview}) that are detected in all coronal AIA channels. This indicates removal of coronal material. In 171\,\AA, the dimmings are pronounced, as majority of the coronal loops in the northern part of the AR (where the eruption occurs) disappear, leaving only rather weak emission. At the location of the slipping motion of the hot channel at 00:40\,UT (denoted by pink circle in Figs. \ref{Fig:Overview}--\ref{Fig:CID_iris}), core dimmings are later observed \citep[downward-pointing white arrow in Fig. \ref{Fig:Overview}b6; see also][for nomenclature of coronal dimmings]{Dissauer2018a,Dissauer2018b}. These core dimmings are located in the interior of the far end of PRH (cf., panel c5); i.e., they likely correspond to the legs of the erupting structure \citep[see, e.g.,][]{Hudson1996,Thompson2000,Gopalswamy2000,Attrill2006,Veronig2021,Krista2022,Wang2023}. Although visible in 171\,\AA, these core dimmings are most pronounced in the 211\,\AA~channel (not shown). However, no CME is ultimately detected in the SOHO/LASCO catalog\footnote{\url{https://cdaw.gsfc.nasa.gov/CME_list/UNIVERSAL_ver1/2020_12/univ2020_12.html}}. Since this event is weak, and located close to the center of the solar disk in a decayed AR, we speculate that the resulting halo CME is also weak and ultimately not detected.

\subsection{Reconnection of the filament with coronal loops\\ (AR--RF reconnection)}
\label{Sect:AR-RF}

We now turn to the analysis of the interaction of the filament with coronal loops and show that this interaction is due to \textit{AR--RF} reconnection \citep[first theoretically described by][]{Aulanier2019} between the original flux rope (filament) and the surrounding coronal loops, producing the hot flux rope (hot channel, Sect.~\ref{Sect:Hot_channel}).
To study the evolution of the coronal loops, we employed the combined improved difference method (CID), which highlights the moving structures and/or changes in the observed intensity. The CID is analogous to the running difference method, but the images are scaled differently to enhance weak structures as well as partially suppress the image noise. Details on the method are given in Appendix \ref{Appendix:CID}.

Since the coronal AIA channels observed both the moving coronal loops, the erupting filament, as well as the flare ribbons, we plot the CID from these observations alongside the CID 304\,\AA, which contains only the erupting filament and flare ribbons, see Fig. \ref{Fig:CID_iris} and the accompanying animation. Since different coronal loop systems are visible in different AIA filters, we plot the CID 211\,\AA~together with the CID 171\,\AA, covering the range of warm coronal loops at the AR periphery. The location of the IRIS FOV (see Sect. \ref{Sect:Coalignment}) is overplotted as the red box, with the central vertical line indicating the IRIS slit. 

The animation associated with Fig. \ref{Fig:CID_iris} shows that the eruption of the filament is accompanied by motions of coronal loops in all coronal channels. The first loop motions are detected around 00:26\,UT, and are clearly induced by the erupting filament. Around 00:30:20\,UT, a multitude of coronal loops (panels a1--b1 of Fig. \ref{Fig:CID_iris}) are seen interacting with the filament, which experiences strong brightenings (white in panel c1). One of the strongest brightenings in the filament is located directly at the IRIS slit (red arrow in panel c1). By 00:40\,UT, all of the filament is in emission (see panel b4 of Fig. \ref{Fig:Overview}).

The interaction of the filament with coronal loops and the associated brightenings are a result of arcade-to-rope reconnection \citep[AR--RF reconnection][]{Aulanier2019}. This can be discerned by following the connectivity of, for example, two points denoted by the orange and pink circles (Figs. \ref{Fig:Overview}--\ref{Fig:CID_iris}). At 00:24, both are outside of PRH and correspond to coronal loops (arcades A1 and A2 in panel b2 of Fig. \ref{Fig:Overview}). Both these coronal loops subsequently disappear, producing only dark traces in the CID(4) images. The A1 disappears by 00:31\,UT (panel a2 of Fig. \ref{Fig:CID_iris}); however, by 00:32\,UT this location is a flux rope (denoted R1 in panel a3 of Fig. \ref{Fig:Overview}), indicating a change of connectivity $A \to R$. This is consistent with the evolution of the hook PRH (Fig. \ref{Fig:ribbon_evolution}: At 00:30\,UT, this location is outside of PRH, indicating that it is a coronal arcade; but at 00:32\,UT, the PRH has already swept it, becoming a part of the flux rope. Still later, at 01:06\,UT, this location is again outside of the PRH, and at the very edge \citep["cantle",][]{Lorincik2021a} of the flare loop arcade (F1 in panel a6). Overall, the evolution of the connectivity is $A \to R \to F$, in good agreement with the multi-stage \textit{AR--RF} reconnection \citep[see Sect. 4.4 and Fig. 6 of][]{Aulanier2019}

The loop A2 disappears later, at 00:36\,UT (panels a5 and b5 of Fig. \ref{Fig:CID_iris}), and at 00:40\,UT, this location is again a flux rope (R2 in panel a4 of Fig. \ref{Fig:Overview}), showing slipping motions of its footpoints. Subsequently, this location is on the inside of the hook PRH (Fig. \ref{Fig:ribbon_evolution}), but the  hot flux rope visible in 94\,\AA~moves away from this location, leaving only a coronal dimming region. 

%
%
%
\section{Spectral analysis}
\label{Sect:Spectra}

\subsection{Rasters}
\label{Sect:Rasters}

The raster mode of IRIS observations allows us to follow the evolution of the flare ribbons and filament interaction with overlying arcades during the eruption. The AR is scanned by 8 slit positions, which provided the spectra in Mg II, Si IV and CII lines within about one minute raster cadence. We provide four  examples of the eight positions of the slit locations in the C II SJI image in Fig.~\ref{Fig:sji_SiIVspectra} left panels, while the right panels show the \ion{Si}{IV} spectra at all 8 slit positions. We choose 4 times where the individual spectra indicate reconnections between filament and coronal loops. These are detected principally at 00:30:20 UT, 00:32:54 UT, 00:34:11 UT, 00:35:56 UT. An accompanying animation provides a complete view of the spectral evolution at the flare site from 00:25:32 to 01:15:28 UT.

\subsection{Spectra of the reconnecting filament}
\label{Sect:Spectra_filament}

By analyzing the spectra at all 8 positions, we found that slit 7 effectively captures the first three reconnection episodes between the overlying arcades and the filament, see blue arrows in Fig. \ref{Fig:sji_SiIVspectra}. The corresponding slit position 7 is shown in Fig.\ref{Fig:CID_iris}, panels (a1)–-(c1), (a3)--(c3), and (a4)--(c4). In all three cases, brightenings are detected in the filament in 304\,\AA~(panels c1, c3, and c4), furthermore at locations where the filament overlaps with the evolving coronal loops. The clearest example occurs at 00:30:20\,UT (panels b1--c1 of Fig. \ref{Fig:CID_iris}), where the 304\,\AA~brightening is the strongest and the unobstructed portion of the 171\,\AA~loop is in close proximity to the slit position.

The spectra of the reconnecting parts of filament, as recorded by the IRIS spectrograph, have first to be distinguished from both the spectra of the flare ribbons in the chromosphere, as well as from the spectra of the erupting filament itself, which will appear to be blue-shifted as it moves upwards in the solar atmosphere. The ribbons are relatively easy to distinguish with the help of the \ion{Mg}{II} spectral window, since the ribbons appear bright also in the extended Mg II line wings. The location of the ribbon is thus clearly indicated by a horizontal emission feature spanning the \ion{Mg}{II} window, see Fig. \ref{Fig:mgii}. The spectral signature of the erupting filament is a large inclined feature crossing both the \ion{Mg}{II} and \ion{Si}{IV} spectra. In \ion{Mg}{II}, the filament is easier to distinguish, since its relatively cool material is seen in absorption. The inclination of the filament in the spectrograph detector images is caused by the twisted structure of the filament, whose southern part is rising faster ($\approx -100$ \kps), while its northern part can be moderately redshifted (up to $\approx +50$ \kps at 00:34:11 UT, panel (i) of Fig. \ref{Fig:mgii}), indicating a combination of upward and untwisting motions during the filament eruption.

The \ion{Si}{IV} and \ion{Mg}{II} spectra from the first three reconnection episodes (at slit position 7) at 00:30:20 UT, 00:32:54 UT, and 00:34:11 UT are shown in detail in Fig. \ref{Fig:mgii}. With the help of the \ion{Mg}{II} lines, it can be immediately seen that the position marked by the dashed blue horizontal line is distinct from the ribbon, whose emission appears further to the north, at all three times. Furthermore, the emission in the \ion{Si}{IV} line at the location of the blue dashed line is both brighter and more blueshifted than in the rest of the filament. The strong blueshifts observed at the three times occur in all three cases in a very localized area of a few IRIS pixels, corresponding to 1--2$\arcsec$. The individual spectra observed at the three times are shown in Fig. \ref{Fig:spectra_} along with their uncertainties. The properties of these spectra are described in the following subsections. Except the presence of strongly blueshifted components, a common feature is relative absence of the \ion{O}{IV} and \ion{S}{IV} lines at 1404.8 and 1406.2\,\AA, which have less than 20 counts (DN) and are barely visible above the continuum. Only the strongest \ion{O}{IV} line at 1401.2\,\AA~is clearly discernible, even if it is weak in some spectra.

\begin{table*}[!h]
\caption{Fit parameters for the IRIS \ion{Si}{IV} FUV spectrum. Individual components plotted in Fig. \ref{Fig:spectra_} are indicated.} 
\label{table1}
\vspace{-0.5cm}
  \begin{center}
  \begin{tabular}{lccccccl}  
  \hline
  \noalign{\smallskip}
    Time        & line          & $I_0$ [DN]      & $\lambda$ [\AA]    & $v_\mathrm{D}$ [\kps]    & FWHM [\AA]         & FWHM [\kps]    & notes \\
  \noalign{\smallskip}
  \hline
  \noalign{\smallskip}
                & \ion{O}{IV}   & ~40.0 $\pm$ 1.5  & 1400.365 $\pm$0.012  & $-$170.8 $\pm$2.7     & 0.706 $\pm$0.062  & 151.1 $\pm$13.3 & Gaussian \\
    00:30:20 UT & \ion{Si}{IV}  & 361.6 $\pm$ 4.6  & 1401.797 $\pm$0.010  & $-$207.9 $\pm$2.2     & 0.567 $\pm$0.023  & 121.2 $\pm$~4.9  & $\kappa$\,=\,3.27 $\pm$0.50 \\
                & \ion{Si}{IV}  & ~79.4 $\pm$ 6.6  & 1402.423 $\pm$0.012  & ~$-$74.2 $\pm$2.5     & 0.429 $\pm$0.227  & ~91.6 $\pm$48.5  & Gaussian \\
                & \ion{Si}{IV}  & ~28.5 $\pm$15.3  & 1402.793 $\pm$0.016  & ~~$+$2.7 $\pm$3.6     & 0.220 $\pm$0.101  & ~47.0 $\pm$21.6  & Gaussian, weak \\
  \noalign{\smallskip}
  \hline
  \noalign{\smallskip}
                & \ion{O}{IV}   & ~22.2 $\pm$ 1.1  & 1400.788 $\pm$0.030  & ~$-$80.3 $\pm$6.3     & 0.891 $\pm$0.122  & 190.6 $\pm$26.1  & Gaussian, weak \\
    00:32:54 UT & \ion{Si}{IV}  & ~64.8 $\pm$ 6.4  & 1401.922 $\pm$0.026  & $-$181.3 $\pm$5.6     & 0.568 $\pm$0.123  & 121.4 $\pm$26.3  & Gaussian \\
                & \ion{Si}{IV}  & 174.8 $\pm$ 3.5  & 1402.457 $\pm$0.014  & ~$-$66.9 $\pm$3.0     & 0.681 $\pm$0.018  & 145.5 $\pm$~3.8  & Gaussian \\
  \noalign{\smallskip}
  \hline
  \noalign{\smallskip}
                & \ion{O}{IV}   & ~13.1 $\pm$ 1.3  & 1400.005 $\pm$0.033  & $-$247.7 $\pm$7.1\tablefootmark{a} & 0.721 $\pm$0.178  & 154.3 $\pm$38.1\tablefootmark{a}& Gaussian, weak \\
    00:34:11 UT &  ?            & ~97.2 $\pm$ 3.0  & 1401.443 $\pm$0.017  & $-$283.1 $\pm$3.6\tablefootmark{b} & 0.789 $\pm$0.080  & 168.6 $\pm$17.1\tablefootmark{b}& Gaussian \\
                & \ion{Si}{IV}  & 203.2 $\pm$ 2.7  & 1402.523 $\pm$0.008  & ~$-$52.8 $\pm$1.8     & 0.820 $\pm$0.083  & 175.2 $\pm$17.7  & Gaussian \\
                & \ion{Si}{IV}  & 239.7 $\pm$ 9.0  & 1403.182 $\pm$0.004  & ~$+$88.1 $\pm$0.9     & 0.404 $\pm$0.016  & ~86.4 $\pm$~3.4                & Gaussian \\
 \noalign{\smallskip}
   \hline
  \end{tabular}
  \end{center}
  \vspace{-0.5cm}
  \tablefoot{
  \tablefoottext{a}{if interpreted as \ion{O}{IV}}
  \tablefoottext{b}{if interpreted as \ion{Si}{IV}}
  }
\end{table*}

\subsubsection{IRIS spectra at 00:30:20\,UT}

The \ion{Si}{IV} and \ion{Mg}{II} spectra observed in the reconnecting filament at 00:30:20\,UT are shown in Fig. \ref{Fig:spectra_}, panels (a) and (d). The \ion{Si}{IV} line is completely blueshifted, with an asymmetric profile. The \ion{O}{IV} line also appears blue-shifted. The spectrum can be fitted with 4 components plus a constant continuum. The overall fit is a satisfactory with $\chi_\mathrm{red}^2$\,=\,1.87 and shown by the orange curve. Although the value of the $\chi_\mathrm{red}^2$ is larger than 1, most of the contributions come from very weak, unfittable features above the continuum \citep[see Appendix B of][]{Dudik2017b} and not from improper fitting of the spectral lines.

The relatively-weak \ion{O}{IV} line can be fitted with a single, strongly blue-shifted Gaussian component with $v_\mathrm{D}$\,=\,$-170.8 \pm2.7$\,\kps (violet dashed line in Fig. \ref{Fig:spectra_}. The \ion{Si}{IV} line requires at least two blue-shifted components, a strong one with $v_\mathrm{D}$\,=\,$-207.9 \pm2.2$\,\kps and a weaker (less intense) one with $v_\mathrm{D}$\,=\,$-74.2 \pm2.5$\,\kps (blue curves). A further weak, slightly redshifted component improves the fit (red Gaussian). The properties of these components are listed in Table \ref{table1}, where $I_0$, $\lambda$, $v_\mathrm{D}$, FWHM, and $\kappa$ denote the intensity, wavelength, doppler velocity, full width half maximum and kappa values respectively. A constant continuum of 10.4\,$\pm$0.3\,DN is sufficient to account for most of the weak spectral features as well as the true continuum (horizontal orange line in Fig.~\ref{Fig:spectra_}).

Interestingly, the strongest \ion{Si}{IV} component is non-Gaussian, with $\kappa$\,=\,3.3\,$\pm$0.5, i.e., with pronounced wings, likely indicating presence of accelerated particles \citep[see the discussion in][]{Jeffrey2016,Jeffrey2017,Dudik2017b,Polito2018}. Its width (FWHM) is also extreme, about $121$\,\kps, indicating presence of MHD turbulence that can also be associated with particle acceleration \citep{Bian2014}. We find that the width of the individual fit components, whether Gaussian or non-Gaussian, are generally rather large (see Table \ref{table1}), being comparable even with widths traditionally detected from flare lines formed above 10\,MK such as \ion{Fe}{XXI} and hotter ones \citep[e.g.,][]{Antonucci1989,Ashfield2024}. Note that the single component used for fitting the \ion{O}{IV} line is usually wider than individual components for the \ion{Si}{IV} line (Table \ref{table1}), while it is also less blue-shifted. This is probably due to the \ion{O}{IV} line being relatively weak (40 DN), so that it cannot be reliably fitted with multiple components like the much stronger \ion{Si}{IV} line. 

Overall, we take the presence of strong dopplershift of the \ion{Si}{IV} line, along with its extreme width, as well as the non-Gaussian profile of the most blue-shifted component as three independent lines of evidence for impulsive energy release due to magnetic reconnection at this location. 
The \ion{Mg}{II} spectra also show the presence of strongly blue-shifted components (see the dashed blue lines in Fig.~\ref{Fig:mgii}, panels g--i). The most blue-shifted peak occurs at about $\approx$200\,\kps (Fig. \ref{Fig:spectra_} d), which corresponds well with the most blue-shifted peak of Si IV (panel a). This strong blue-shift in \ion{Mg}{II} is also quite localized, as the spectra more to the north (above the blue dashed line in Fig.~\ref{Fig:mgii}) the spectra exhibit a dark-reddish area in the core of the line. This area is more extended in panels h and i and corresponds to the escaping filament. 

To better appreciate these \ion{Mg}{II} spectra with strong Doppler-shifts (Fig. \ref{Fig:spectra_}), we overplotted the chromospheric spectra for reference (green lines in Fig. \ref{Fig:spectra_}). The reference  spectra are taken in the far northern part of the IRIS FOV, at the pixel No. 700, far away from the erupting filament. The reference spectrum is characterized by the presence of a narrow central reversal, and much narrower widths.
These \ion{Mg}{II} spectra could be interpreted by a cloud model consisting of two clouds: a low-velocity blue-shifted absorbing cloud representing the erupting filament (around $-$40\,\kps) and a strongly blue-shifted (around $-$200\,\kps) emission cloud that originates due to reconnection similarly as the correspondingly strong blue-shifted \ion{Si}{IV} components. Note that these \ion{Mg}{II} profiles are also similar to the profiles observed in a previous study of reconnection jet \citep{Joshi2020FR,Joshi2021}.

Interestingly, unlike at later times, at 00:30:20\,UT the far \ion{Mg}{II} wings of the reconnecting filament are higher than the reference profile, possibly indicating overall heating of the filament material.

\subsubsection{IRIS spectra at 00:32:54\,UT}

At 00:32:54\,UT, the filament plasma at slit position 7 is again bright and showing strong blueshifts. The \ion{Si}{IV} line is asymmetric but fully blue-shifted. Consequently, it can be fitted with 2 components (panel b of Fig. \ref{Fig:spectra_}). A weaker but strongly blue-shifted component is found in the asymmetrically enhanced blue wing, having $v_\mathrm{D}$\,=\,$-181.3$\,$\pm$5.6\,\kps. The main component is stronger, but less blue-shifted ($v_\mathrm{D}$\,=\,$-66.9$\,$\pm$\,3.0\,\kps, see Table \ref{table1}). Both these components are again very broad but Gaussian, having FWHMs of 121 and 146\,\kps, respectively. 

The \ion{Mg}{II} lines are also quite broad (especially compared to the reference profile), with no central reversal, and have a pronounced, asymmetric blue wing at $\approx$150\,\kps, in broad agreement with the \ion{Si}{IV} spectra. The far wings (above about 300\,\kps) are similar to  the reference profile wings (green color).
This non-reversed \ion{Mg}{II} profile in Fig.~\ref{Fig:spectra_} (panel e) is  the result of filament absorption, as evident in the spectra shown in Fig.~\ref{Fig:mgii} (panel h). 
In comparison to the \ion{Si}{IV} profiles, the \ion{Mg}{II} profiles are significantly affected by filament absorption. Again at  00:32:54 UT  along the blue dashed line
 (panel i), the emission is reduced all along the \ion{Mg}{II} blue wing due to the smeared distribution of the filament velocities during its eruption.
\subsubsection{IRIS spectra at 00:34:11\,UT}

At 00:34:11\,UT, the \ion{Si}{IV} spectrum becomes rather complex with several distinct components. Each of these can be fitted with a single Gaussian to a good degree of accuracy. The most intense component is a red-shifted one with $v_\mathrm{D}$\,=\,$+$88.1\,$\pm$0.9\,\kps (Fig. \ref{Fig:spectra_} and Table \ref{table1}). There is also a relatively strong blue-shifted component with $v_\mathrm{D}$\,=\,$-$52.8\,$\pm$1.8\,\kps.

There are two other lines, one moderately strong at 1401.44\,\AA; i.e., in the vicinity of the rest wavelength of the \ion{O}{IV} 1401.16\,\AA~line, and a weak line at 1400.01\,\AA. What these lines are is not immediately clear. The weak line is most likely an \ion{O}{IV} one, either a strongly blue-shifted ($v_\mathrm{D}$\,=\,$-247.7$\,$\pm$7.1\,\kps) component of the line whose rest wavelength is at 1401.16\,\AA, or a red-shifted ($v_\mathrm{D}$\,=\,$+50.3$\,$\pm$7.1\,\kps) component of a weak \ion{O}{IV} line whose rest wavelength is at 1399.77\,\AA~\citep[see Table 1 in][]{Dudik2014}. The moderately strong line, denoted by the "?" sign in panel (c) of Fig. \ref{Fig:spectra_}, can either be interpreted as a distinct \ion{Si}{IV} component with $v_\mathrm{D}$\,=\,$-283.1$\,$\pm$3.6\,\kps or a red-shifted \ion{O}{IV} one with $v_\mathrm{D}$\,=\,$+59.9$\,$\pm$3.6\,\kps. We think that the blue-shifted \ion{Si}{IV} interpretation is more likely, since if this was the red-shifted \ion{O}{IV} component, the comparatively strong blue-shifted counterpart of this line would be missing. Further support that this an extremely blue-shifted \ion{Si}{IV} component comes from the \ion{Mg}{II} spectra, which show a weak but broad blue-shifted feature at about $\approx$250\,\kps (pink arrow in panel (f) of Fig. \ref{Fig:spectra_}). 

In accordance with the multiple blue- and red-shifted components being present in the \ion{Si}{IV} line, the \ion{Mg}{II} spectrum also shows presence of multiple components with quite similar velocities. Furthermore, both the \ion{Mg}{II} as well as \ion{Si}{IV} lines are again extremely broad, with the blue-shifted components typically being the broadest ones (Table \ref{table1}). The non-thermal width of the two blue-shifted \ion{Si}{IV} components is 175 and 168\,\kps, respectively.

The very weak (~15 DN) signal in the Mg II line and the corresponding line component in the Si IV window (97 DN) with high blueshift (~$-250$ \kps) likely also arise due to reconnection. However, due to the presence of multiple overlapping structures along the line of sight, it is difficult to isolate a distinct source for these highly blueshifted components in either AIA or IRIS observations (see Fig.~\ref{Fig:CID_iris}, panels a4–c4). These faint signals may originate from different structures entirely and could also be explained by Doppler dimming effects \citep{Rompolt1980,  Peat2024}.

%
%
\section{Discussion}
\label{Sect:Discussion}

MHD simulations of solar eruptions predict that reconnection in 3D happens within the volume of the QSLs, where the field lines can exchange connectivities, and not at a rather singular `X'-type location as in the 2D picture. The 3D nature of magnetic reconnection subsequently manifests itself in the slipping motion of the reconnecting structures, including the erupting flux rope \citep{Aulanier2012,Janvier2013,Aulanier2019}. Indeed, we have found that the drifting footpoints of the reconnecting flux rope undergoing $AR-RF$ reconnection are slipping (see Sect. \ref{Sect:Hot_channel} and panel (a4) of Fig. \ref{Fig:Overview}). 

Nevertheless, our spectroscopic observations show strong, but rather localized response to the $AR-RF$ reconnection within the erupting filament, that is, within the original erupting flux rope. This localization is rather strict, to within 1--2$\arcsec$ (Sect. \ref{Sect:Spectra_filament}, and also restricted in time to several episodes. We note that even though this localization in space and time is rather imperfect on account of the IRIS being a rastering spectrometer, with the raster cadence of about one minute in our case it is unlikely that IRIS would miss a large portion of the reconnection-induced dynamic changes within the filament plasma. 

The plasma response to reconnection was found to be threefold. First, we observe rather large blue-shifts (up to 200\,\kps or more) in both the \ion{Si}{IV} and \ion{Mg}{II} lines. Second, these blue-shifts are always accompanied by very large non-thermal widths of the \ion{Si}{IV} line, above 100\,\kps. Third, in one instance at the beginning of the $AR-RF$ reconnection, the strongest and most blue-shifted \ion{Si}{IV} line is also non-Gaussian, being best approximated by a rather low value of $\kappa$\,$\approx$3.3. Such non-Gaussian profiles were in flares detected previously only in much hotter lines, such as \ion{Fe}{XVI} \citep{Jeffrey2016,Jeffrey2017} and \ion{Fe}{XXIII}--\ion{Fe}{XXIV} \citep{Jeffrey2017,Polito2018}, whose formation temperature (in equilibrium conditions) is orders of magnitude higher.

The large blue-shifts have been observed previously during reconnection between jet and loops \citep{Ruan2019} or during breakout events \citep{Reeves2015}. Our observations presented here mean that these spectral signatures extend also to the $AR-RF$ reconnection geometry.

The large non-thermal widths, along with the strong blue-shifts and brightenings in the erupting filament, likely indicate that the filament plasma is being heated by the reconnection. Supporting evidence for this can be found in the AIA imaging observations. First, portions of the filament appear in emission even in coronal filters such as 171\,\AA, 193\,\AA, and 211\,\AA~(cf., Fig. \ref{Fig:CID_iris}). Second, the hot emission in \ion{Fe}{XVIII} appears at about 00:30\,UT (see animation accompanying Fig. \ref{Fig:Overview}), that is, at the time IRIS observed the first spectral signatures of reconnection. Subsequently, the hot flux rope is already well-developed by 00:32\,UT, and the flare arcade also starts to appear, as shown by the panel (a3) of Fig. \ref{Fig:Overview}. Except perhaps for the presence of \ion{Fe}{XVIII}, spectroscopic observations of the reconnecting filament in lines formed at temperatures of 1--10\,MK would be required for direct confirmation of the rapid heating of the filament plasma to much higher temperatures, as the original flux rope turns to flare loop ($R \to F$) and coronal loop arcades to new flux rope field lines ($A \to R$).

One last question remains: Why do we observe mostly strong blue-shifts? The traditional reconnection scenario in an `X'-point (which is not entirely valid here for reasons explained above) expects the presence of bi-directional reconnection outflow jets. In our case, the filament reconnects with coronal loops that are already significantly hotter than the formation temperatures for the IRIS \ion{Mg}{II} and \ion{Si}{IV} lines. Therefore, the strongly red-shifted counterparts to our observed blue-shifts again likely occur at much higher temperatures.

%
%
%
\section{Summary}
\label{Sect:Summary}

We present spectroscopic analysis of a filament reconnecting with coronal loops during the course of its eruption that was accompanied by a weak B3.2-class flare that occurred on 2020 December 9. Despite this event being relatively small, we observe strong response of the filament plasma to reconnection, chiefly in terms of $\approx$200\,\kps blue-shifts and extremely large widths of the \ion{Si}{IV} lines as observed by IRIS.

This accompanying flare is a classical two-ribbon event, with both ribbons showing J-shaped hooks indicating the presence of an erupting magnetic flux rope. This flux rope is first identified as the erupting filament and later a hot flux rope seen in the \ion{Fe}{XVIII} line in the 94\,\AA~passband of SDO/AIA. One of the ribbons is compact, located in relatively-strong magnetic polarities of the AR. The conjugate ribbon is weak and extended, and its hook undergoes significant evolution, sweeping conjugate footpoints of several coronal loops. These coronal loops initially overlap with the erupting filament along the line of sight. The eruption of the filament leads to interaction with the coronal loops, which subsequently reconnect with the filament and disappear. The filament turns to a hot flux rope whose footpoints reach to the previous footpoints of the coronal loops, indicating that the erupting filament underwent significant heating as it reconnected in the arcade-to-rope $AR-RF$ reconnection geometry \citep{Aulanier2019}.

The IRIS slit rastered the erupting filament in the location where it interacted with the coronal loops. Using careful co-alignment of SDO/AIA and IRIS, we identified locations of the individual ribbons, brightenings of the erupting filament, and coronal loops along the line of sight. The IRIS spectra shows the erupting filament as slanted structure in absorption in \ion{Mg}{II} and in emission in \ion{Si}{IV}, while the ribbon is visible as a bright structure in the \ion{Mg}{II} wings. On top of that, we distinguished the spectral signatures of the reconnecting filament, which are threefold:
\begin{itemize}
    \item Presence of strong blue-shifts, up to $\approx$200\,\kps, and higher in some cases, as distinct components of the \ion{Si}{IV} and \ion{Mg}{II} lines,
    \item presence of extremely large non-thermal widths, routinely exceeding 100\,\kps in the \ion{Si}{IV} line,
    \item presence of a strongly non-Gaussian profile ($\kappa$\,$\approx$\,3.3) of the most blue-shifted component at the start of the reconnection.
\end{itemize}
Overall, these three separate lines of evidence show strong response by the filament plasma to the energy being released via to the $AR-RF$ reconnection with the neighboring coronal loop arcade. Although the traditional bilateral flows are not seen in the reconnecting filament except in the last instance, we argue that the red-shifted component is not seen by IRIS because the filament is reconnecting with much hotter coronal loops, so that the red-shifted counterpart cannot be seen in the relatively-cool lines of \ion{Mg}{II} and \ion{Si}{IV} observed by IRIS.

In summary, our observation of the response of the filament plasma to reconnection presents first spectral observations of the $AR-RF$ reconnection, revealing that this response can be quite localized within the reconnecting filament.



%
\begin{acknowledgements}
This research has been supported by the Research Council of Norway through its Centres of Excellence scheme, project number 262622 and  the European Research Council through the Synergy grant No. 810218 (“The Whole Sun,” ERC-2018-SyG). 
Part of this work was supported by the German \emph{Deut\-sche For\-schungs\-ge\-mein\-schaft, DFG\/} project number Ts~17/2--1. J.D. acknowledges support from the Czech Science Foundation, grants No. GACR 22-07155S and 25-18282S, as well as institutional support RWO:67985815 from the Czech Academy of Sciences. 
G.A. acknowledges financial support from the French national space agency (CNES), as well as from the Programme National Soleil Terre (PNST) of the CNRS/INSU also co-funded by CNES and CEA. R.C. acknowledges the support from DST/SERB project No. EEQ/2023/000214. 
H$\alpha$ Data were acquired by GONG instruments operated by NISP/NSO/AURA/NSF with contribution from NOAA. AIA data are courtesy of NASA/SDO and the AIA, EVE, and HMI science teams. IRIS is a NASA small explorer mission developed and operated by LMSAL with mission operations executed at NASA Ames Research Center and major contributions to downlink communications funded by ESA and the Norwegian Space Centre. 
\end{acknowledgements}

\bibliography{2024_Eruption_2020-12-09_arXiv}
\bibliographystyle{aa}

\clearpage
\appendix

\section{Combined Improved Difference of SDO/AIA images}
\label{Appendix:CID}

To study the evolution of various structures in time, we employ the "combined improved difference" (hereafter, CID) that combines the advantages of both the traditionally used running-difference method and the less common log running ratio method.

The running difference (RD) of a series of imaging observations is simply constructed as
\begin{equation}
    \mathrm{RD}(d;x,y)_i = I(x,y)_i - I(x,y)_{i-d}\,,
    \label{Eq:RDIFF}
\end{equation}
where $x$ and $y$ are spatial coordinates and $i$ denotes the $i$-th image in a time series. The $d$ gives the difference between frames chosen. In practice, $d$ is often larger than one for observations with sufficient cadence to cover the evolution of individual structures. For example, \citet{Dudik2017a} chosen $d$\,=\,2 (i.e., a difference of 24\,s) to highlight the evolution of coronal loops in the vicinity of eruptive solar flares observed by SDO/AIA. Meanwhile, \citet{Dudik2019} used $d$\,=\,4 to better highlight coronal loops involved in \textit{AR--RF} reconnections. We note that a higher value of $d$ generally produces a more contrasting signal in the RD series, as long as $d$ is still smaller than the typical evolution timescale of the structure. 

The log running ratio method \citep[LRR,][]{Lorincik2021b,Lorincik2024} is also sometimes used to highlight dynamical evolution. Its expression is 
\begin{equation}
    \mathrm{LRR}(d;x,y)_i = \log_{10}\left(\frac{I(x,y)_i}{I(x,y)_{i-d}}\right)\,,
    \label{Eq:LRR1}
\end{equation}
which can be rewritten as
\begin{equation}
    \mathrm{LRR}(d;x,y)_i = \log_{10}(I(x,y)_i) - \log_{10}(I(x,y)_{i-d})\,,
    \label{Eq:LRR}
\end{equation}
meaning that LRR is essentially a running difference of logarithmically-scaled images. This makes it useful for studying evolution in environments containing multiple structures of highly different intensities, where the linear RD method leads to some structures structures either being difficult to detect and different ones saturated. 

\begin{figure}[t]
   \centering
\includegraphics[width=0.5\textwidth]{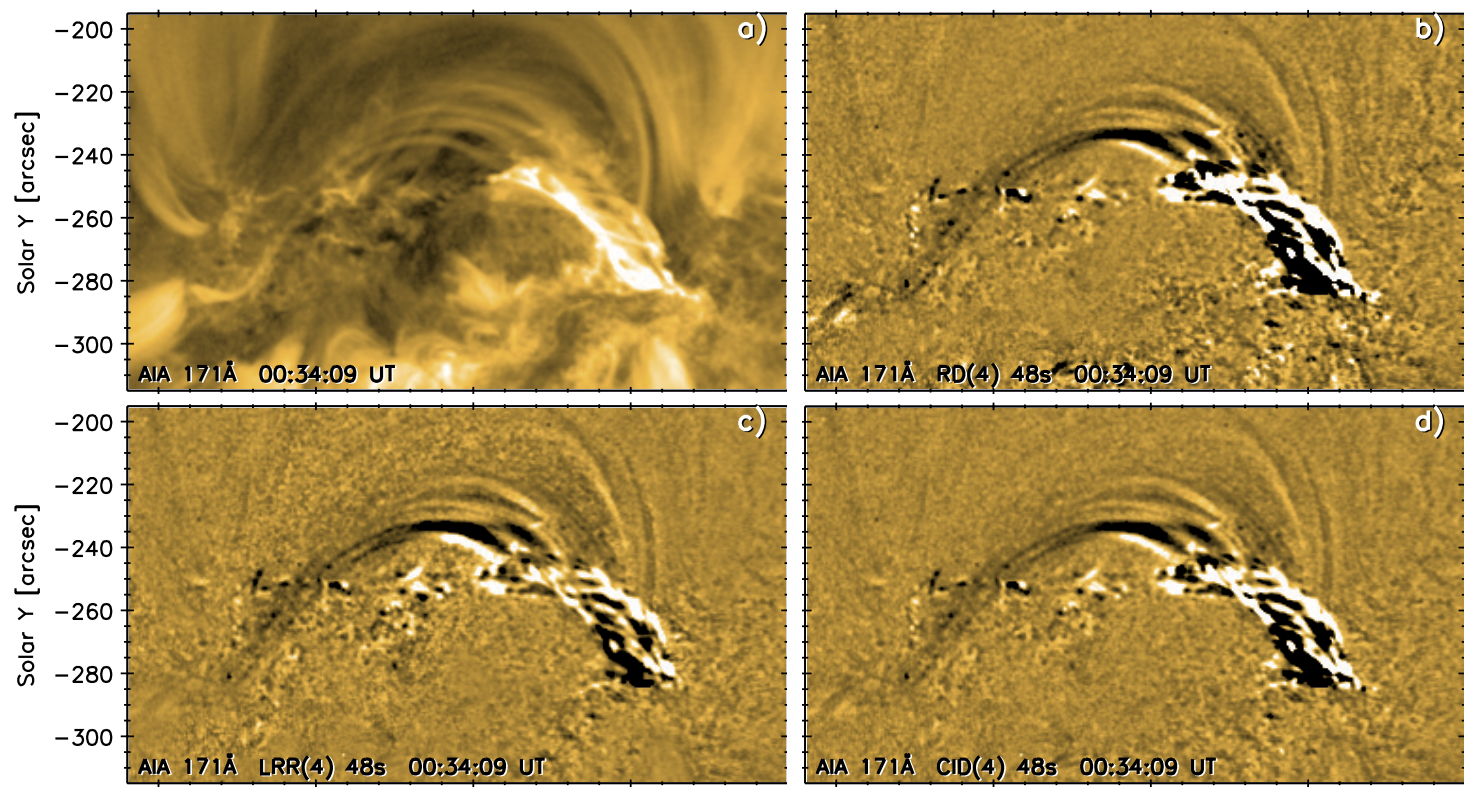}   
\caption{Comparison of three difference methods. Panel (a) shows the original AIA 171\,\AA~observations, while panels (b)--(d) show the running difference RD(4), log running ratio LRR(4) and combined improved difference CID(4), calculated using a time difference of $d$\,=\,4.}
   \label{Fig:CID_method}
\end{figure}

Both RD and LRR methods, while useful, contain their own drawbacks. The RD method enhances noise in areas of medium observed intensity but non-moving or weakly moving structures. An example for AIA 171\,\AA~at 00:34:09\,UT can be found in panel (b) of Fig. \ref{Fig:CID_method}. Contrary to that, the LRR greatly enhances noise in weak-signal areas (panel c), where the photon and/or read noise is larger or comparable to the real signal. This in turn could necessitate averaging either in the image plane (for example, by a 3$\times$3 boxcar) or in time, in each case leading to a slight loss of resolution.

Given that the noise-enhanced areas are in practice complementary in the RD and LRR methods, their product can lead to suppression of both areas of enhanced noise, while preserving real difference signal originating in moving or evolving structures Upon trial and error, we found best results for the following scaling, which we call the "combined improved difference" (CID):
\begin{equation}
    \mathrm{CID}(d)_i = \lvert \mathrm{RD}(d)_i \rvert^{1/2} \times  \lvert\mathrm{LRR}(d)_i\rvert^{1/2} \times \frac{\mathrm{RD}(d)_i}{\lvert\mathrm{RD}(d)_i\rvert}\,,
    \label{Eq:CID}
\end{equation}
where we omitted the dependence on $(x,y)$ for the sake of brevity. Note the last multiplicator ensures the sign of CID$(d)_i$ is the same as that of RD$(d)_i$ and LRR$(d)_i$. The resulting CID(4), shown in panel (d) of Fig. \ref{Fig:CID_method}, produces clear and contrasted images, and retains all evolving large-scale structures. Based on examination of our event, we found that the CID is especially useful for AIA filters such as 171\,\AA, 193\,\AA, 211\,\AA, 304\,\AA, and areas of higher signal in 131\,\AA~and 335\,\AA. For the 94\,\AA, which in our event contains only relatively weak signals, the LRR leads to strong enhancement of noise. For this filter and the present event, we therefore find both LRR and CID inferior to RD. 

%
\section{\ion{Fe}{XVIII} proxy from SDO/AIA}
\label{Appendix:Fe18}

To isolate the hot emission of flare loops, we use the method of \citet{DelZanna2013multith}, where the intensity of \ion{Fe}{XVIII} is given by
\begin{equation}
    I(\ion{Fe}{XVIII}) = I(94\,\AA) - I(211\,\AA)/120 - I(171\,\AA)/450\,,
    \label{Eq:Fe18}
\end{equation}
where the coefficients were determined by \citet{DelZanna2013multith} mostly for observations taken in 2010, that is, relatively close to the AIA launch date. However, EUV instruments are known to degrade in sensitivity with time \citep[e.g.,][]{BenMoussa2013}, and AIA is no exception \citep{Boerner2014}. Since its launch, the AIA instrument has degraded significantly \citep[][]{DosSantos2021}. Particularly affected are the longer-wavelength channels of 211\,\AA, 335\,\AA, and 304\,\AA, while the shorter-wavelength channels less affected. To correct the degradation of the AIA sensitivity, we use the version 10 of the AIA effective areas, available from SolarSoft. We find that the sensitivity of the 94\,\AA~and 171\,\AA~channels have decreased to a factors of about 0.86 and 0.72 relatively to pre-launch sensitivities. Contrary to that, the relative sensitivity of the 211\,\AA~channel on 2020 December 9 is only about 0.44. To calculate the \ion{Fe}{XVIII} proxy, the equation (\ref{Eq:Fe18}) is corrected here for decrease of sensitivity of individual filters.

\end{document}